\def\wordcount {1}		

\ifdefined\preprint
  \documentclass[preprint,review,11pt]{elsarticle}
\fi
\ifdefined\wordcount
  \documentclass[final,3p,times,twocolumn]{elsarticle}
\fi
\ifdefined\final
  \documentclass[final,3p,times,twocolumn]{elsarticle}
\fi

\usepackage{graphicx,stfloats,amsmath,inputenc}
\usepackage{color}

\biboptions{sort&compress}

\newcommand{\peh}[1]{\textcolor{black}{#1}}

\journal{Proceedings of the Combustion Institute}

\begin{document}

\begin{frontmatter}

\title{Particle Pair Dispersion and Eddy Diffusivity in a High-Speed Premixed Flame}

\author[fir]{Ryan Darragh}

\author[sec]{Colin A.Z. Towery}
\author[thi,fou]{Alexei Y. Poludnenko}
\author[sec]{Peter E. Hamlington\corref{cor1}}
\ead{peh@colorado.edu}

\address[fir]{Department of Aerospace Engineering Sciences, University of Colorado, Boulder, 80309, USA}
\address[sec]{Department of Mechanical Engineering, University of Colorado, Boulder, 80309, USA}
\address[thi]{Department of Mechanical Engineering, University of Connecticut, Storrs, 06269, USA}
\address[fou]{Department of Aerospace Engineering, Texas A\&M University, College Station, TX 77843}
\cortext[cor1]{Corresponding author:}

\begin{abstract}

Turbulent mixing is a physical process of fundamental importance in high-speed premixed flames. This mixing results in enhanced transport of temperature and chemical scalars, leading to potentially large changes in flame structure and dynamics. To understand turbulent mixing in non-reacting flows, a number of classical theories have been proposed to describe the scaling and statistics of dispersing fluid particle pairs, including predictions of the effective, or turbulent, eddy diffusivity. Here we examine the validity of these classical theories through the study of fluid particle pair dispersion and eddy diffusivity in highly turbulent premixed methane-air flames at a Karlovitz number of approximately $140$. Using data from a direct numerical simulation and a higher-order Lagrangian tracking algorithm, particle pair centroids are seeded at different initial temperatures and separations, and then integrated forward in time. We show that scaling relations and results developed for pair dispersion in non-reacting flows remain relevant in this high-intensity premixed flame, and we identify the impacts of heat release on dispersion and eddy diffusivity.

\end{abstract}


\end{frontmatter}

\ifdefined \wordcount
\clearpage
\fi

\section{Introduction} \label{sec:intro}

Turbulence leads to substantially enhanced rates of scalar mixing in a wide range of natural and engineering flows. This enhanced mixing is especially important in turbulent premixed combustion, since the properties and evolution of scalar quantities such as temperature and fuel mass fraction have leading-order effects on the rate of fuel consumption and heat release. Moreover, in many high-speed practical applications such as scramjet and detonation engines, the turbulence intensity is high and turbulent mixing can substantially alter the structure of the flame, leading to, for example, flame broadening, distributed burning, and broken reaction zones \cite{driscoll2019}. 

Although turbulent mixing is fundamentally an advective process whereby large turbulence-induced scalar gradients lead to more rapid molecular transport, this overall process is often referred to as simply ``turbulent diffusion.'' Beginning with the foundational work by \citet{richardson1926}, and recognizing the fundamental connection to turbulent advection, a common approach to studying turbulent diffusion has been to examine the dispersion, or separation, of pairs of fluid particles. In cases where particles separate more rapidly, turbulent diffusion is stronger, and \citet{richardson1926} predicted that the turbulent diffusivity is related to the 4/3 power of the distance between the two particles. This relation has since been extensively studied and validated in a variety of non-reacting turbulent flows~\cite{falkovich2001,sawford2001,salazar2009,scatamacchia2012,malik2018,biferale2014,bitane2012}.

Substantially less focus has been placed on the connection between fluid particle dispersion and turbulent diffusivity in reacting flows. \citet{chaudhuri2015} examined the dispersion of pairs of flame (as opposed to fluid) particles during hydrogen-air premixed combustion, finding that the early-time behavior of the particle pairs is non-universal and varies based on the temperature of the isosurface on which the flame particles reside. For the fluid particles that are the focus of the present study, however, this dependence on flame location may not be as strong, particularly at high turbulence intensities. Indeed, other tests of classical turbulence theories (such as the Kolmogorov hypotheses and corresponding scaling laws) have shown that turbulence properties during high-speed premixed combustion are generally similar to those found in non-reacting flows~(e.g., \cite{hamlington2011,bobbitt2016a,whitman2019}). 

In the present study, we seek to understand whether heat release affects pair dispersion and turbulent diffusion in highly turbulent premixed flames, and whether classical theories from non-reacting turbulence remain applicable for such conditions. This study will address three questions in particular: (i) Are classical scaling laws and relations derived for non-reacting turbulent dispersion applicable in high-speed premixed combustion? (ii) How does heat release affect particle dispersion and the applicability of classical scaling laws and relations? (iii) How does turbulent diffusion vary at different locations within a premixed flame?

We address these questions by tracking pairs of fluid particles in a direct numerical simulation (DNS) of a premixed methane-air flame with a Karlovitz number (Ka) of roughly 140. In the following, we provide background on the study of particle pair dispersion and describe the DNS and analysis. We then show that theories developed for pair dispersion in non-reacting flows are applicable in this high-intensity premixed flame, with relatively limited impacts of heat release on dispersion and the turbulent effective eddy diffusivity.

\section{Background} \label{sec:background}
Fluid particle pairs consist of two infinitesimal parcels of fluid whose motions are determined by the local instantaneous flow velocity. The locations of each particle are denoted $\boldsymbol{x}^{(1)}(t)$ and $\boldsymbol{x}^{(2)}(t)$, where $t=0$ corresponds to the time at which particle tracking begins (or when the particles are released). The instantaneous displacement between the two particles is $\boldsymbol{D}(t) = \boldsymbol{x}^{(2)}(t) - \boldsymbol{x}^{(1)}(t)$, where the initial displacement is $\boldsymbol{D}_{0}=\boldsymbol{D}(0)$ and the initial separation distance is $D_0 = \left|\boldsymbol{D}_0\right|$. The relative displacement of a single particle is $\boldsymbol{R} = \boldsymbol{D}-\boldsymbol{D}_0$, and the mean-square relative displacement is $\langle R^2\rangle = \langle|\boldsymbol{D}(t) - \boldsymbol{D}_{0}|^2\rangle$, where $R=|\boldsymbol{R}|$ and $\langle\cdot\rangle$ indicates an average over all tracked pairs in a flow. In the present analysis of premixed flames, this average is taken over all particle pairs released with a centroid at a particular value of the temperature.

The temporal scaling of $\langle R^2\rangle$ changes as particle pairs evolve and separate. At very small times, particles evolve ballistically and at intermediate times they enter the Richardson range, which is analogous to the inertial range identified by Kolmogorov \cite{monin1975}. At large times, the particles in a pair become decorrelated and independent of their initial separation, thus entering the diffusive range described by Taylor~\cite{taylor1922}. Each of these ranges are outlined in the following sections.

\subsection{Ballistic range}
At very small times, particle velocities are roughly constant~\cite{batchelor1952}. If $D_0 \gg\eta$, where $\eta=(\nu^3/\varepsilon)^{1/4}$ is the Kolmogorov length scale given kinematic viscosity $\nu$ and the average rate of turbulence kinetic energy dissipation $\varepsilon$, then this will hold for times $0\le t \ll t_0$, where $t_0  = D_0^{2/3}/\varepsilon^{1/3}$ is the Batchelor time, defined as the eddy turnover time at the length scale of the initial particle separation. If $D_0\ll \eta$, then the velocities will be roughly constant in the range $0 \le t \ll t_{\eta}$~\cite{monin1975}, where $t_\eta = (\nu/\varepsilon)^{1/2}$ is the Kolmogorov time. While a pair of particles is in this range, their displacement can be approximated by a Taylor expansion about $t=0$ as~\cite{polanco2018}
\begin{equation}
    \boldsymbol{D}(t) = \boldsymbol{D}_0 + \delta \boldsymbol{v}_0 t + \frac{1}{2}\delta \boldsymbol{a}_0t^2 + \mathcal{O}\big(t^3\big)\,,
\end{equation}
where $\delta \boldsymbol{v}_0 = \partial \boldsymbol{D}(t)/\partial t|_{t=0}$ and $\delta \boldsymbol{a}_0 = \partial^2 \boldsymbol{D}(t)/\partial t^2|_{t=0}$ are, respectively, the relative velocity and acceleration at $t=0$. The mean-square relative displacement in the ballistic range is thus given by
\begin{equation}
    \big\langle R^2\big\rangle(t) = \left<\delta \boldsymbol{v}_0\cdot \delta \boldsymbol{v}_0\right> t^2 + \left<\delta \boldsymbol{v}_0\cdot\delta \boldsymbol{a}_0\right>t^3 + \mathcal{O}\big(t^4\big).
\end{equation}
For very small times, the leading order term is dominant, and the $t^2$ dependence of $\langle R^2\rangle$ is a characteristic feature of particles separating in the ballistic range.

\subsection{Richardson range}
If the initial separation of the particles is much less than the energy injection scale, eddies larger than $D_0$ will advect the particles together, changing their position in space but not the relative distance between the pair. By the locality hypothesis \cite{richardson1926}, only eddies of a size similar to the separation distance are expected to be effective at changing the relative distance between particles. If, additionally, $D_0 \gg \eta$ and $t$ is large enough such that initial conditions have been forgotten, or if $D_0\ll \eta$ and the later time  separation distance is in the inertial range, Kolmogorov's second hypothesis of scale similarity applies, assuming a large enough Reynolds number~\cite{monin1975}. Consequently, it can be assumed that the statistics will depend only on the length scale and $\varepsilon$. By taking the time-dependent distance between two particles, $D(t)=|\boldsymbol{D}(t)|$, as the length scale and combining it with $\varepsilon$, the turbulent diffusivity is given as \cite{richardson1926}
\begin{equation}\label{eq:K}
    K\big(\big\langle D^2\big\rangle\big) = k_0\varepsilon^{1/3}\big\langle D^2\big\rangle^{2/3}\,,
\end{equation}
where $k_0$ is a dimensionless constant. This relation is known as the four-thirds law and is attributed to Richardson, who arrived at it experimentally, and to Obhukhov, who derived it using similar arguments to those presented here~\cite{richardson1926,obukhov1941,monin1975}. This law is analogous to the Kolmogorov four-fifths law for the scaling of third order structure functions \cite{monin1975} and states that the diffusivity increases with separation distance.

The scaling of $\langle R^2\rangle(t)$ in the Richardson range can be obtained by combining Eq.\ \eqref{eq:K} with Einstein's definition of fluid particle diffusivity, given as \cite{einstein1905}
\begin{equation}\label{eq:Einst}
K\big(\big\langle D^2\big\rangle\big) = \frac{1}{6}\frac{d\big\langle D^2\big\rangle}{dt}\,.
\end{equation}
Integrating this definition using Eq.\ \eqref{eq:K} yields
\begin{equation}\label{eq:Richg}
    \big\langle R^2\big\rangle(t) = g\varepsilon t^3\,,
\end{equation}
where $g$ is the dimensionless Richardson constant. Although $g$ has traditionally been difficult to measure precisely~\cite{sawford2001}, recent estimates have ranged from roughly 0.5 to 0.6~\cite{jullien1999,ott2000,boffetta2002,biferale2005,buaria2015,sawford2008}. In order to accurately determine $g$, there must be a large separation of scales, thus requiring a sufficiently high Reynolds number in the flow.

In addition to the predicted $t^3$ scaling of $\langle R^2\rangle$ in the Richardson range, both \citet{richardson1926} and \citet{batchelor1952} provided analytical predictions for the distribution of relative separation distances $R$. In particular, Richardson predicted that the probability density function (pdf) of $R' = R/\langle R^2\rangle^{1/2}$, denoted $p(R',t)$, should obey the isotropic diffusion equation~\cite{monin1975,richardson1926}
\begin{equation}
    \frac{\partial p(R',t)}{\partial t} = \frac{1}{R'^2}\frac{\partial}{\partial R'}\left[R'^2K(R',t)\frac{\partial p(R',t)}{\partial R'}\right]\,,
\end{equation}
which can be solved~\cite{buaria2015} as
\begin{equation}\label{eq:RichP}
    p(R',t) = \alpha R'^2\exp\big(-\beta R'^{2/3}\big)\,,
\end{equation}
where $\alpha=117$ and $\beta=5.44$. By contrast, Batchelor predicted a Gaussian distribution for $p(R',t)$ by arguing that integration over the velocity amounts to an application of the central limit theorem~\cite{batchelor1952}. In practice, for non-reacting flows, the Richardson solution in Eq.\ \eqref{eq:RichP} has been observed at intermediate times, and a Gaussian-like distribution has been observed at later times outside the Richardson range~\cite{sawford2001,schumacher2008}.

\subsection{Diffusive range}
At very long times, the particles in each pair become decorrelated and spread according to single particle statistics. Practically, \citet{taylor1922} showed that, in this ``diffusive'' range, the \peh{mean-square} separation between two particles increases linearly with $t$, and there is no dependence on the initial separation. Here we track particle pairs over sufficiently long durations to capture the scaling of $\langle R^2\rangle$ in both the ballistic and Richardson ranges, but only begin to approach this diffusive range scaling. However, it will be seen that the Gaussian-like prediction for $p(R',t)$ from Batchelor is approximately recovered at long times in high-intensity turbulent premixed flames, consistent with observations in non-reacting turbulence~\cite{schumacher2008}.

\section{Direct Numerical Simulation} \label{sec:num_sim}
A direct numerical simulation (DNS) of premixed methane-air combustion has been performed by solving the compressible reactive-flow Navier-Stokes equations~\cite{towery2016} using \verb Athena-RFX ~\cite{stone2008,poludnenko2010}. This code solves the governing equations on a fixed, equispaced, three-dimensional (3D) mesh using the unsplit corner transport upwind scheme~\cite{colella1990,saltzman1994} with a nonlinear HLLC Riemann solver and piecewise-parabolic spatial reconstruction. Periodic forcing at the scale of the domain width, $L$, is performed throughout the entire domain and has been confirmed to produce statistically stationary turbulent flames~\cite{poludnenko2010}. Overall, the simulation has third-order spatial and second-order temporal accuracy. A 19-step reduced chemical mechanism \cite{vie2015} is used to model stoichiometric methane-air combustion at atmospheric conditions. All physical parameters used to initialize and setup the simulations are shown in Table~\ref{tab:disp_flame_setup}.

The computational domain consists of an unconfined prismatic box with an aspect ratio of $1\times1\times16$, with periodic boundaries in all three directions prior to ignition; this configuration has been described in numerous previous studies (e.g., \cite{poludnenko2010,hamlington2011,hamlington2017,whitman2019}). Turbulence develops in the domain for one large-scale eddy turnover time $\tau_L$ (see Table \ref{tab:disp_flame_setup}) before a planar laminar flame with a normal direction along the $z$ axis is introduced near the center of the domain. After ignition, the $z$ boundary conditions are switched to be zero-order extrapolation, allowing flow into and out of the domain, while the $x$ and $y$ boundaries remain periodic. 

The simulation domain is discretized using $256\times 256\times 4096$ grid cells, giving a resolution of $2.7\times 10^{-3}$~cm. This provides half a grid cell per unburnt Kolmogorov scale, $\eta_\mathrm{r}$, and six grid cells per burnt Kolmogorov scale, $\eta_\mathrm{p}$ (see Table \ref{tab:disp_flame_setup}). \peh{This also corresponds to 16 grid cells per laminar flame thermal width $\delta_\mathrm{f}$, which was found in previous studies \cite{poludnenko2010} to be sufficient for resolving the flame when thin radical regions are largely absent, as is the case in the present methane-air flame.} The simulation is run for one $\tau_L$ prior to ignition and is then run for another one $\tau_L$ prior to starting data collection. At this point, data are output at a high frequency to allow the calculation of Lagrangian trajectories; the data collection occurs over roughly $5\tau_L$.

\begin{table}[t!]
    \centering
    \small
    \begin{tabular}{lll}
\hline
$L$				& $0.70037$ cm			& Domain width \\
$T_\mathrm{r}$		& $300$ K					& Unburnt temperature \\
$P_\mathrm{r}$		& $1.01325\times10^6$ erg/cm$^3$	& Unburnt pressure \\
$\rho_\mathrm{r}$	& $1.12\times 10^{-3}$ g/cm$^3$	& Unburnt density \\
$\delta_\mathrm{f}$	& $4.38\times 10^{-2}$ cm	& Laminar \peh{thermal} width \\
$S_\mathrm{f}$		& $37.2$ cm/s				& Laminar flame speed \\ \hline
$\ell$			& $2.05\times 10^{-1}$ cm	    & Unburnt integral scale \\
$U'_\ell$			& $785$ cm/s				& Unburnt integral vel.\ \\
$\tau_L$			& $5.922 \times 10^{-4}$ s	& Eddy turnover time \\
$Da$		& $0.19$					& Damk\"{o}hler number \\
$Ka$		& $142$					& Karlovitz number \\ \hline
$\varepsilon$		& $1.181\times 10^9$ erg/g$\cdot$s	& Energy dissipation rate  \\
$\eta_\mathrm{r}$			& $1.382\times 10^{-3}$ cm	& Unburnt Kolm.\ length \\
$\eta_\mathrm{p}$			& $1.734\times 10^{-2}$ cm	& Burnt Kolm.\ length \\
$t_{\eta_\mathrm{r}}$		& $1.174\times 10^{-5}$ s		& Unburnt Kolm.\ time  \\
$t_{\eta_\mathrm{p}}$		& $6.340\times 10^{-5}$ s		& Burnt Kolm.\ time \\
$Re_{T_\mathrm{r}}$			& 87						& Unburnt Reyn.\ number \\
$Re_{T_\mathrm{p}}$		& 28						& Burnt Reyn.\ number\\ 
\hline
    \end{tabular}
    \caption{Physical model parameters of the premixed methane-air flame DNS examined in the present study.\label{tab:disp_flame_setup}}
\end{table}

Using the 3D volumes of velocity output by the DNS, fluid particles are tracked using a Lagrangian algorithm \cite{darragh2017,hamlington2017}. Each particle is evolved in time using 4th and 5th order Runge-Kutta time integration to check accuracy; the 5th-order results are used in the present analysis. Spatial and temporal interpolations are performed using Akima splines, giving at least 2nd order accuracy.

\peh{The centroids of particle pairs are seeded uniformly in spanwise (i.e., $x$-$y$) planes at specific values of the temperature along the $z$ direction. The particle pairs are centered at 300, 600, 1000, 1400, 1800, and 2100~K, giving results that span the unburnt reactants, flame region, and burnt products, as well as within the preheat zone where mixing is strongest.} At each centroid, three pairs of particles are seeded in the $x$, $y$, and $z$ directions, with initial pair separations of $D_0/\eta_\mathrm{r} =$ 1/8, 1/4, 1, 4, 16, and 32. \peh{In total, we examine 65,536 particle pairs at each temperature for each initial separation, resulting in over 14 million trajectories in the present study.}

\peh{Each particle pair is tracked for $5\tau_L$, corresponding to roughly three thermal-width flame crossing times $t_\mathrm{f}=\delta_\mathrm{f}/S_\mathrm{f}=1.2$~ms. The analysis is, however, focused almost entirely on times less than $t_\mathrm{f}$, suggesting that the present results are indicative of particle behavior within the flame. In particular, $t_{\eta_\mathrm{p}}$, corresponding to the approximate end of the ballistic range, is nearly 20 times smaller than $t_\mathrm{f}$, and the integral time scale $t_\ell$, roughly corresponding to the end of the Richardson range, is nearly 1.5 times smaller. As a result, only 0.23\% of particles with initial centroids at 300~K reached a temperature of 1800~K (roughly corresponding to the end of the region of peak heat release) by $t_{\eta_\mathrm{p}}$ and 28\% of these particles reached this temperature by $t_\ell$. For the higher centroid temperature of 1400~K, only 37\% of particles reached 1800~K by $t_{\eta_\mathrm{p}}$, with 81\% by $t_\ell$. Thus, many of the particles remain within the flame for the duration of the ballistic and Richardson ranges, both of which end before the characteristic flame crossing time $t_\mathrm{f}$.}

\section{Results and Discussion} \label{sec:results}
\subsection{Particle dispersion scaling and statistics}
\begin{figure}[t!]
	\centering
	\includegraphics[scale=1]{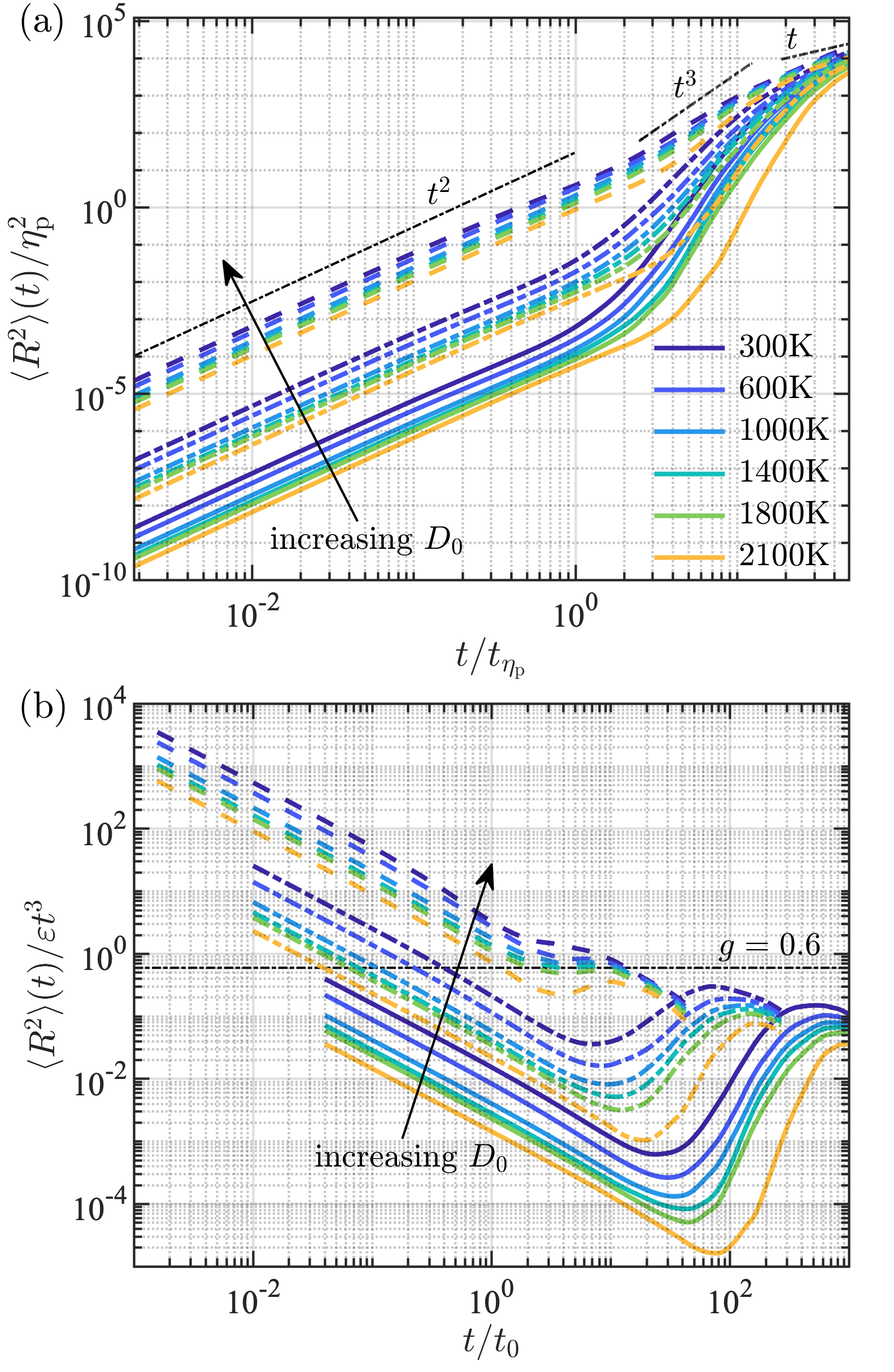}\vspace{-0.15in}
	\caption{Uncompensated (a) and compensated (b) time series of the mean-square relative separation $\langle R^2\rangle(t)$ for particle pairs initially separated in the $z$ direction with centroids at six different temperatures. Results are shown for initial displacements $D_0/\eta_\mathrm{r}=1/8$ (solid lines), 1 (dash-dot lines) and 16 (dashed lines). The horizontal dash-dot line in (b) indicates the estimated value of the Richardson constant reported by \citet{sawford2008}, $g=0.6$.}
	\label{fig:figure1}
\end{figure}

The time evolution of $\langle R^2\rangle$ for particle pairs released with an initial separation in the $z$ direction is shown in Fig.~\ref{fig:figure1}. Results for initial separations in the $x$ and $y$ directions are similar and are not shown here. In Fig.~\ref{fig:figure1}(a), $\langle R^2\rangle$ is normalized by $\eta_\mathrm{p}$, and time is normalized by $t_{\eta_\mathrm{p}}$, both taken in the products (see Table \ref{tab:disp_flame_setup}). At small times, $\langle R^2\rangle$ for each initial separation exhibits a $t^2$ scaling, as expected in the ballistic range, with relatively small differences in the time evolution based on the temperature. \peh{There is, however, an offset in $\langle R^2\rangle$ that depends on temperature, with smaller values of $\langle R^2\rangle$ occuring for higher temperatures. This is most likely due to differences in the temperature-dependent local viscosity at the particle centroids, while the subsequent evolution of $\langle R^2\rangle$ is largely independent of temperature.}

At intermediate times, the displacement speed begins to increase, as indicated by an increasing slope of $\langle R^2\rangle$ in Fig.~\ref{fig:figure1}(a). Of the initial separations shown, only those with the largest $D_0$ appear to approach the $t^3$ scaling predicted in the Richardson range. Pairs with a smaller $D_0$ achieve a greater than $t^3$ scaling due to the relatively low Reynolds number of the present case and the resulting influences from the ballistic (i.e., short time) and diffusive (i.e., long time) ranges. Within this range, however, there is relatively little observed temperature dependence beyond that which may be attributed to variations in the temperature-dependent viscosity. At large times, displacements approach linear scaling with $t$ as particle pairs become decorrelated in the diffusive range and lose their dependence on the initial separation.

The compensated mean-square relative displacements, $\langle R^2\rangle/\varepsilon t^3$, shown as a function of $t/t_0$ in Fig.~\ref{fig:figure1}(b), indicate the time period over which the Richardson range scaling is present. Here we use a constant value of $\varepsilon$ for all separations and temperatures, since it has been shown that $\varepsilon$ is only weakly dependent on temperature in high-Karlovitz number flames \cite{whitman2019}. The theory from \citet{richardson1926} states that, if $D_0$ for a pair of particles is in the inertial range, then the corresponding compensated displacement will decrease linearly until leveling off at a plateau. This linear decrease is expected from the ballistic scaling \cite{batchelor1952}, and the plateau corresponds to the scaling from Eq.~\eqref{eq:Richg} in the Richardson range. 

Both the linear decrease and the plateau are evident in Fig.~\ref{fig:figure1}(b) for $D_0/\eta_\mathrm{r} = 16$. By contrast, $\langle R^2\rangle/\varepsilon t^3$ for particle pairs with smaller $D_0$ falls below the plateau before approaching it from below. This is because the separation distance must be in the inertial range to be on the plateau, and since the separation is increasing it will eventually be in the inertial range. In order to hit the plateau, however, the separation distance must be in the inertial range after the particles have forgotten their initial conditions; i.e., after the max of $t_\eta$ and $t_0$, but before they become affected by large-scale effects. Thus, if the particles do not separate fast enough, they may never approach the plateau, as shown in Fig.~\ref{fig:figure1}(b) for the pairs with smaller initial separations.

\begin{figure}[t!]
	\centering
	\includegraphics[scale=1]{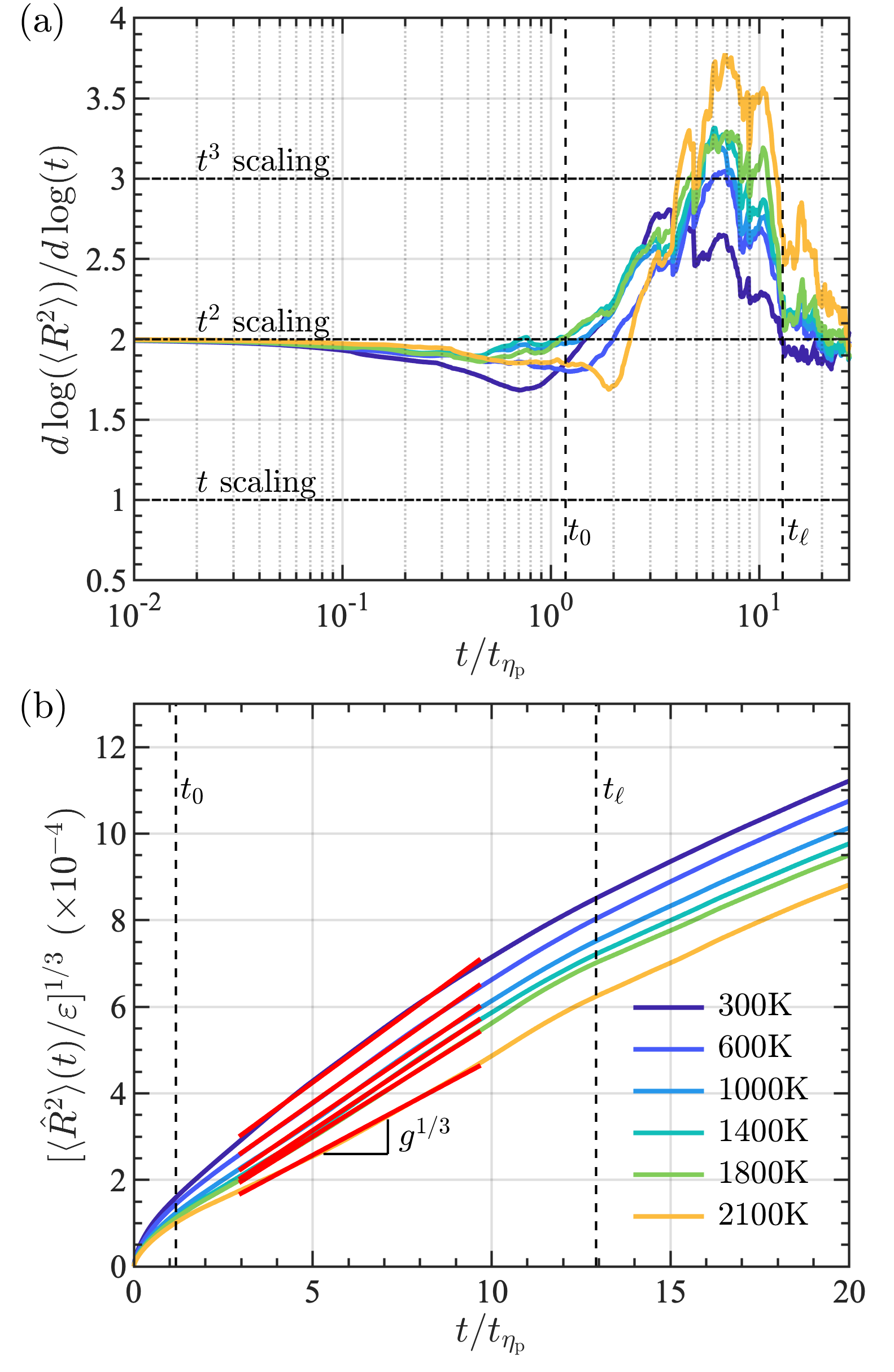}\vspace{-0.15in}
	\caption{Local scaling exponent of $\langle R^2\rangle(t)$ from $d\log(\langle R^2\rangle)/d\log(t)$ for $D_0/\eta_\mathrm{r} = 16$ (a). Panel (b) shows local linear fits of $\big[\big\langle\hat{R}^2\big\rangle(t)/\varepsilon\big]^{1/3}$ (red lines), giving the Richardson constant as $g^{1/3}$.}
	\label{fig:figure2}
\end{figure}

The range over which the $t^3$ scaling of $\langle R^2\rangle$ applies is further indicated by the derivatives of $\log(\langle R^2\rangle)$ with respect to $\log(t)$ for $D_0/\eta_\mathrm{r}=16$, shown in Fig.~\ref{fig:figure2}(a). For $t<t_{\eta_\mathrm{p}}$, $\langle R^2\rangle$ at all temperatures scales as $t^2$. Between $t_0$ and $t_\ell$, by contrast, $\langle R^2\rangle$ for all temperatures approaches the Richardson range scaling of $t^3$. Due to the relatively low Reynolds number of the present case, the range over which this $t^3$ scaling applies is relatively small, and there does seem to be a temperature dependence of the scaling exponent. In particular, the intermediate temperature results more completely reach the $t^3$ scaling, while the lowest and highest temperatures have scaling exponents below and above three, respectively. For $t>t_\ell$, the scaling exponents decrease as the particle pairs enter the diffusive range.

\begin{figure}[tb]
	\centering
	\includegraphics[scale=1]{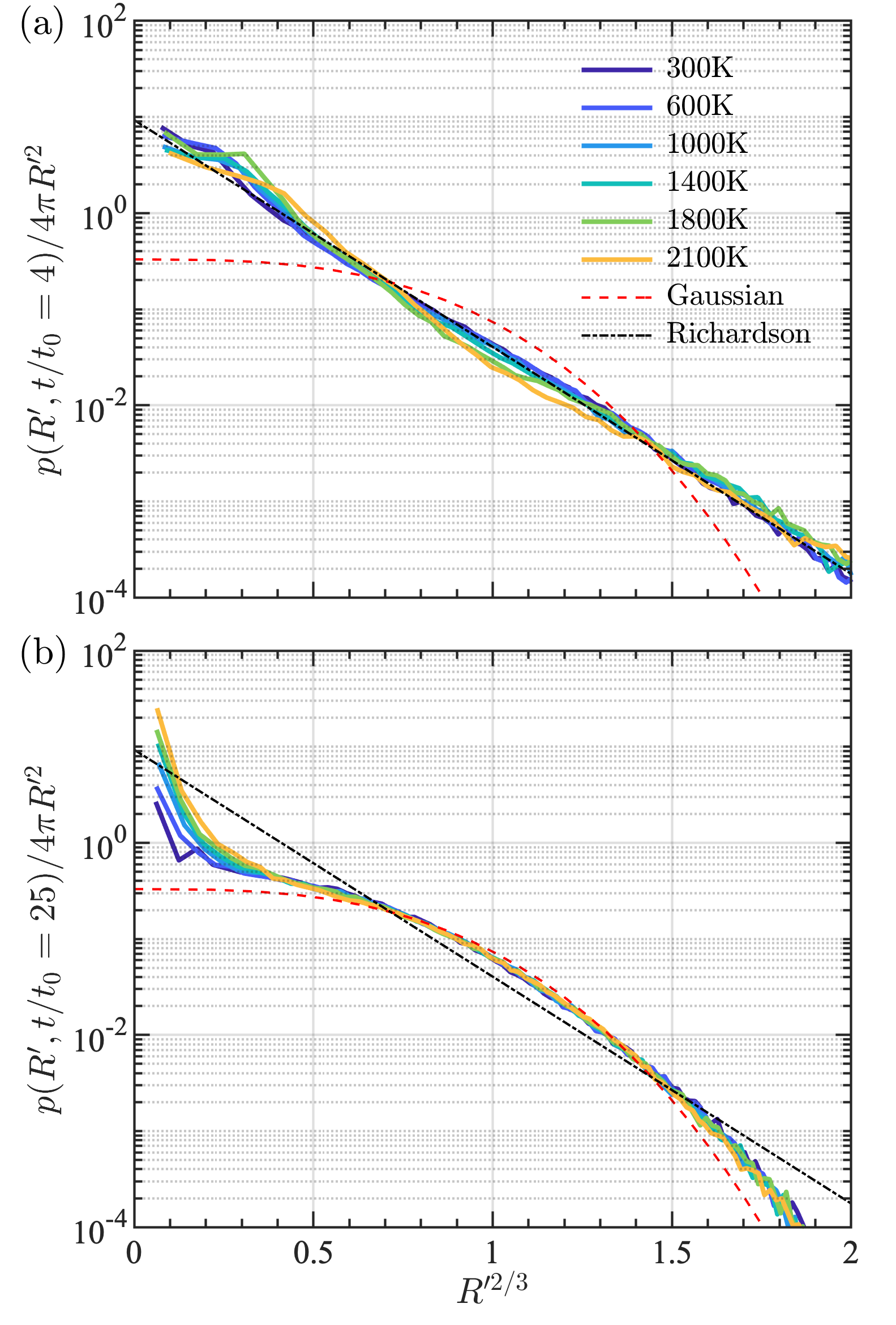}\vspace{-0.15in}
	\caption{Probability density functions (pdfs) of $R' = R/\langle R^2\rangle^{1/2}$ for $D_0/\eta_\mathrm{r}=16$ at (a) $t/t_{0}=4$ and (b) $t/t_0=25$. Gaussian distributions and the Richardson prediction from Eq.~\eqref{eq:RichP} are also shown.}
	\label{fig:figure3}
\end{figure}

The Richardson constant $g$ can be estimated from the curves of $\big[\big\langle\hat{R}^2\big\rangle(t)/\varepsilon\big]^{1/3}$ shown in Fig.~\ref{fig:figure2}(b), where $\hat{R} = |\boldsymbol{D}(t)-\boldsymbol{D}_0 - \delta\boldsymbol{v}_{0}t|$, as a function of $t$ within the Richardson range~\cite{ott2000,sawford2008,buaria2015} for $D_0/\eta_\mathrm{r}=16$. The subtraction of the relative velocity $\delta\boldsymbol{v}_{0}t$ is required to reduce the impact of the ballistic range scaling on the estimate of $g$ (for further details, see \citep{sawford2008}). Given the relation in Eq.~\eqref{eq:Richg}, the slopes of these curves, shown in Fig.~\ref{fig:figure2}(b), give $g^{1/3}$. \peh{Estimates of $g$ for initial centroid temperatures within the flame (i.e., for $1000$~K, $1400$~K, and $1800$~K) are between 0.55 and 0.70, bracketing the classical non-reacting value $g=0.6$ estimated by \citet{sawford2008}. There is some temperature dependence of this constant, however, with $g$ decreasing from a maximum value of $0.88$ at 300~K to $0.34$ at 2100~K. This indicates that, within the Richardson range for large $D_0$, the scaling relation in Eq.~\eqref{eq:Richg} is valid in high-intensity turbulent premixed flames, but that the Richardson constant is temperature-dependent, generally decreasing with increasing temperature.}

The pdfs of $R'$ in Fig.~\ref{fig:figure3} provide a test of the Richardson prediction in Eq.~\eqref{eq:RichP} and the Gaussian-like prediction from Batchelor. At intermediate times in Fig.~\ref{fig:figure3}(a), the pdfs for each temperature are in good agreement with the Richardson prediction. The pdfs at time $t/t_{0}=4$ shown in Fig.~\ref{fig:figure3}(a) correspond to the $t^3$ scaling range indicated in Fig.~\ref{fig:figure2}, further suggesting that we do capture a relatively short Richardson range in these simulations, despite the low Reynolds number. 

By contrast, the pdfs at $t/t_{0}=25$ in Fig.~\ref{fig:figure3}(b) are in much closer agreement with the Gaussian-like pdf. This corresponds to times greater than $t_\ell$ where we would expect diffusive range behavior, and the recovery of the Gaussian-like pdf in Fig.~\ref{fig:figure3}(b) is consistent with prior results at long times in non-reacting turbulence \cite{sawford2008,buaria2015}. It should be noted that there is no apparent temperature dependence of the pdfs in Fig.~\ref{fig:figure3} at either time, and the results from Richardson and Batchelor thus appear to be independent of the location in the flame.

\begin{figure}[t!]
	\centering
	\includegraphics[scale=1]{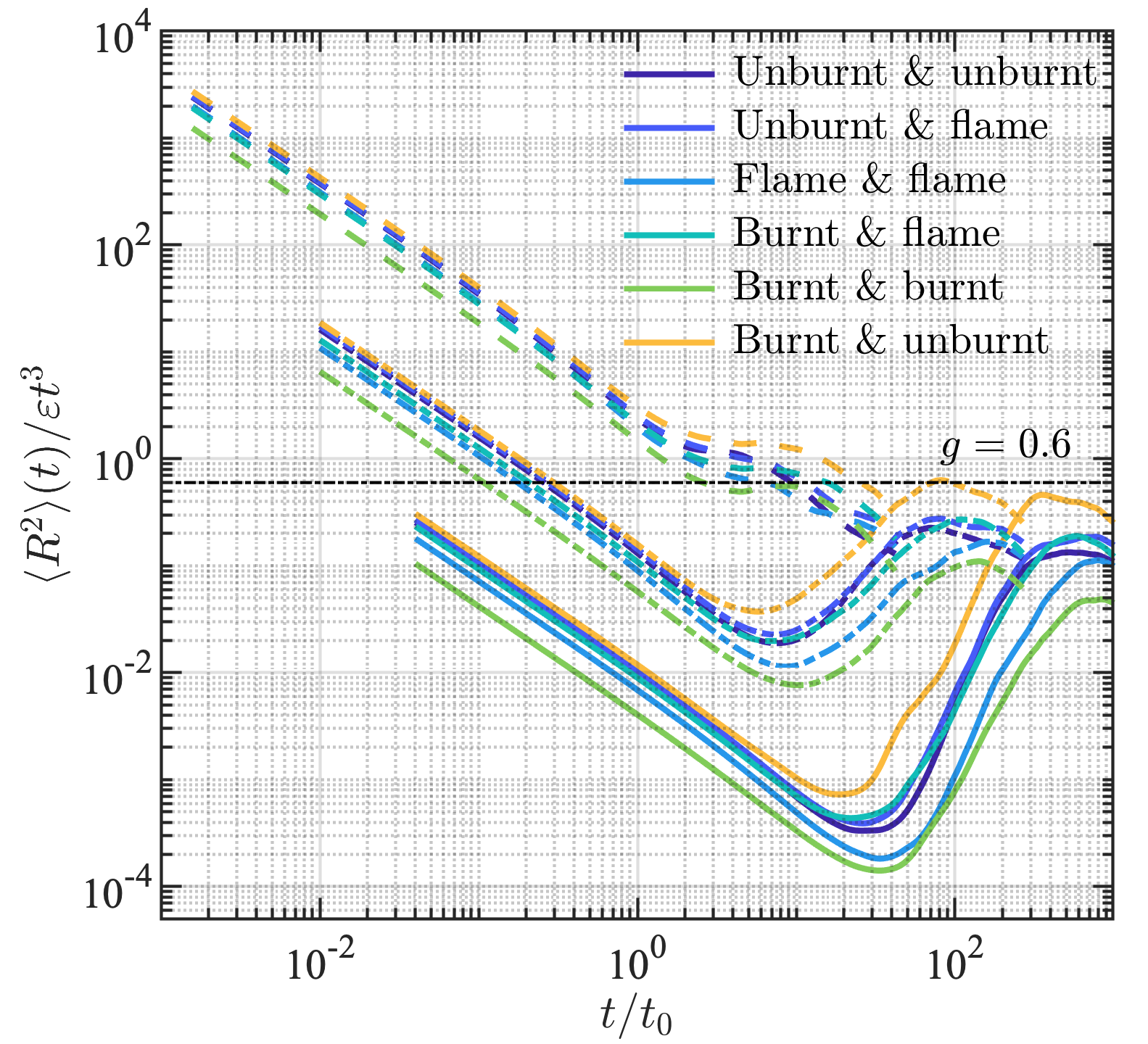}\vspace{-0.15in}
\caption{Compensated time series of $\langle R^2\rangle(t)$ conditioned on the location of each particle at $t=t_\mathrm{f}$ in either the unburnt reactants ($T<500$~K), flame region ($500~\mathrm{K}\leq T\leq1800~\mathrm{K}$), or burnt products ($T>1800$~K). Results are shown for initial displacements $D_0/\eta_\mathrm{r}=1/8$ (solid lines), 1 (dash-dot lines) and 16 (dashed lines).}
	\label{fig:figure5}
\end{figure}

\peh{Finally, we note that these results are specific to the particular type of conditioning used in this analysis. That is, we examine the dispersion and subsequent evolution of fluid particles beginning at the same location within the premixed flame. However, after initialization, the particles are free to then separate such that either particle may re-enter the unburnt reactants, proceed to the burnt products, or remain within the flame. Figure~\ref{fig:figure5} shows results for the compensated mean-square relative displacements, $\langle R^2\rangle/\varepsilon t^3$, depending on whether each particle was in the unburnt reactants (defined as $T<500$~K), burnt products ($T>1800$~K), or within the flame ($500~\mathrm{K}\leq T\leq1800~\mathrm{K}$) at $t=t_\mathrm{f}$. In general, these results are largely consistent with the results shown in Fig.~\ref{fig:figure1}(b), revealing a dependence on the final locations of the particles, although there is still a general correspondence to classical scaling laws in the ballistic, Richardson, and diffusive ranges. Further conditioning of these results is left as a direction for future research.}

\subsection{Turbulent diffusivity}
\begin{figure}[t!]
	\centering
	\includegraphics[scale=1]{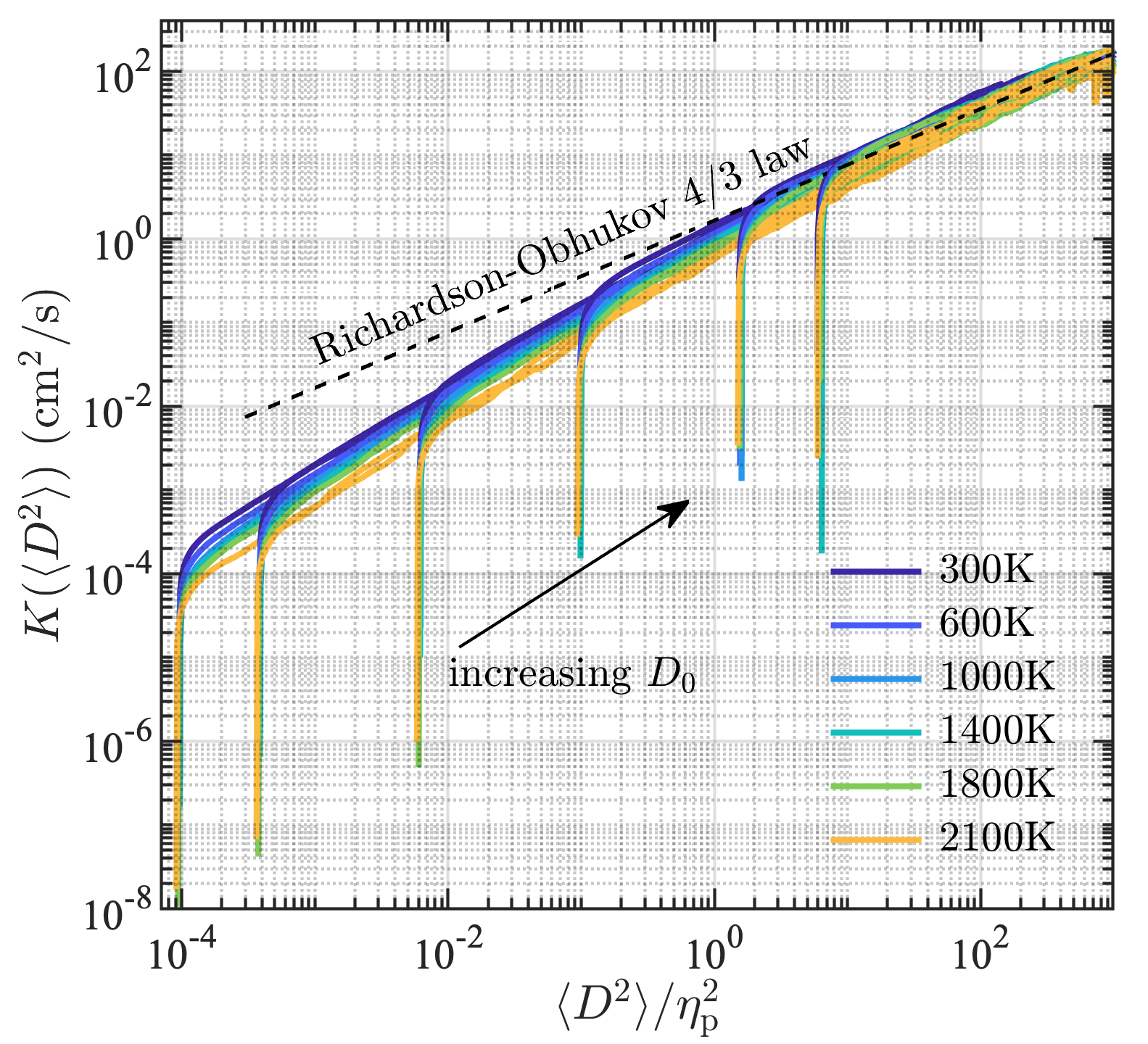}\vspace{-0.15in}
	\caption{Turbulent diffusivity $K(\langle D^2 \rangle)$ for each initial separation and centroid temperature. Diffusivities for increasing $D_0$ increase along the direction indicated by the arrow. The dashed black line corresponds to the Richardson-Obhukov 4/3 law in Eq.~\eqref{eq:K}.}
	\label{fig:figure4}
\end{figure}

The effective eddy, or turbulent, diffusivity is computed from the calculated values of $\langle D^2 \rangle(t)$ for each temperature and initial separation using Eq.\ \eqref{eq:Einst}. Figure~\ref{fig:figure4} shows that $K(\langle D^2 \rangle)$ increases with $\langle D^2 \rangle$ and approaches the Richardson-Obhukov four-thirds scaling law from Eq.~\eqref{eq:K} for large $\langle D^2 \rangle$. \peh{Significantly, results for all temperatures and initial separations approach this scaling law, although there is a persistent difference in the magnitude of $K$ with temperature. In particular, $K$ is largest for small temperatures, corresponding to locations within the preheat zone of the flame where mixing is strongest, and smallest for large temperatures, corresponding to locations in the fully burnt products.}

\peh{From a modeling perspective, Fig.\ \ref{fig:figure4} suggests that the scaling relation between particle pair dispersion and turbulent diffusivity is independent of location in the flame for the highly-turbulent case examined here, but the magnitude of the diffusivity does vary through the flame. This indicates that there may be predictable and potentially universal aspects of turbulent mixing in high-speed combustion. However, further studies spanning a broader range of flow configurations and turbulence conditions must be explored before predictability or universality can be fully established.}

\section{Conclusions} \label{sec:conclusions}
Using data from a DNS of a highly-turbulent methane-air premixed flame and a Lagrangian analysis, we have examined the dispersion of fluid particle pairs in high-speed premixed combustion. In general, scaling laws and statistical relations in the ballistic, Richardson, and diffusive ranges developed for non-reacting pair dispersion are found to remain largely valid in the highly-turbulent premixed flame examined here. In particular, despite the relatively low Reynolds number of the present case, we observe $t^3$ scaling of the mean-square relative separation $\langle R^2 \rangle$ and a correspondence with the Richardson pdf for particle displacements in Eq.~\eqref{eq:RichP}; both of these results indicate the recovery of the Richardson range in the simulations. 

The Richardson constant was estimated using the cubic local slope approach and found to have a value of $g\!\approx\!0.6$ \peh{for intermediate temperatures within the flame}, in agreement with previous non-reacting studies at higher Reynolds numbers. \peh{For lower and higher temperatures, however, the Richardson constant was found to be larger and smaller, respectively, than the classical value.} Finally, turbulence diffusivity was found to approach the Richardson-Obhukov law in Eq.~\eqref{eq:K} for all initial temperatures and separations, \peh{although the magnitude of the diffusivity was temperature dependent.}

Overall, this study suggests that many aspects of fluid particle dispersion and turbulent diffusivity are largely similar to non-reacting results for turbulent premixed flames at high turbulence intensities. \peh{We do observe a dependence on temperature and, hence, location in the flame, but many of these differences may be due solely to changes in the local viscosity and dissipation rate, both of which are temperature dependent.} This suggests that classical non-reacting theories of turbulence and models for turbulent mixing may be relevant at such highly-turbulent conditions. Future research is, however, required to expand the generality of these results for other fuels and flame configurations, \peh{including realistic flow configurations where mean shear is present,} as well as to examine the applicability of non-reacting results as the turbulence intensity decreases. 

\section*{Acknowledgments} \label{sec:acknowledgements}
RD was supported by a National Defense Science and Engineering Graduate Fellowship. CAZT and PEH were supported, in part, by AFOSR award FA9550-17-1-0144 and NSF award 1847111.

\end{document}